\documentstyle[amssymb,preprint,aps]{revtex}

\begin{document}
\title{The quantum conductance of ballistic microconstrictions in metals with an
open Fermi surface.}
\author{A. Namiranian$^{\left( 1\right) }$, Yu.A. Kolesnichenko$^{\left( 2\right) }$}
\address{$^{\left( 1\right) }${\small \ }${}${\small \ Institute for Advanced Studies}%
\\
{\small in}\\
{\small Basic Sciences, }\\
{\small 45195-159, Gava Zang, Zanjan, Iran }\\
$^{\left( 2\right) }${\small \ B.Verkin Institute for Low Temperature}\\
{\small Physics and Engineering, }\\
{\small National Academy of Sciences of Ukraine, }\\
{\small 47 Lenin Ave., 310164 Kharkov, Ukraine}}
\date{\today}
\maketitle

\begin{abstract}
It is shown that the conductance $G$ of the quantum microconstriction in the
metal with an opened Fermi surface as a function of the contact diameter
undergoes the jumps $e^{2}/h$ of the opposite sign. The negative jumps is
the result of the limitation of the energy of the electron motion along the
direction, in which the Fermi surface is opened. The point contact spectrum $%
dG/dV$ of such constriction has additional peaks at the bias $eV$ when the
maximal energy $\varepsilon _{\max }$ of the quantum subband is equal to the
energies $\varepsilon _{F}\pm \frac{eV}{2}$ ($\varepsilon _{F}$ is the Fermi
energy).
\end{abstract}

The quantum size effect in conductors was theoretically predicted by
I.Lifshits and A.Kosevich in 1955 \cite{LifKos} and experimentally it was
found in thin films of metals and semiconductors (see, for example, \cite
{Komnik}). In these studies the quasiclassical oscillation of
thermodynamical and kinetic properties have been investigated, because to
limited range of the sample thicknesses $d$, which were then accessible (as
usual $d\gg \lambda _{F},$ $\lambda _{F}$ is the Fermi wave length). The
advances in the modern technology of nanofabrication make it possible to
realize the ultraquantum limit in the size effect in conducting properties
by using the small ballistic contacts, which size is comparable to the Fermi
wave length. The current $I$ through such microconstriction defines by
currents of one dimensional quantum subbands, each of them contribute to the
conductance $G=dI/dV$ value $G_{0}=2e^{2}/h$ ($V$ is the voltage applied to
the contact). As a result, the conductance $G$ displays a step-like
structure versus contacts size. For ''large'' contacts ($d\gg \lambda _{F}$)
this structure turns to the quasiclassical oscillations. Firstly the effect
of the conductance quantization was observed in a model system in the
two-dimensional (2D) electron gas formed at a $GaAs-Al_{x}Ga_{1-x}As$
heterojunction \cite{2Dfirst,2Dfirst-too}. The development of methods
of the scanning tunnel microscopy and mechanically controllable break
junctions enables to investigate the conductivity of ultrasmall (up to
atomic size) contact in real metals \cite{STM1,STM2,BJ1,BJ2}.
 By using these methods the quantum steps of conductance were observed
in three dimensional (3D) point contacts. In the simple metals $\left(
Na,Cu,Au\right) $ the steps of the conductance rather similar to the
conductance of 2D contacts \cite{BJ2,Krans,Krans2}. But for
metals with a more complicated electronic structure, such as $Al$ and $Pt,$
the size dependence of the conductance has a more irregular behavior, as
compared to the simple metals \cite{Krans}. In $Al$ and $Pt$ the first few
conductance plateaus have positive slope. It signifies that the conductance
decreases when the contact size increases. In some cases the slightly
pronounced negative steps of the conductance had been observed \cite{BJ2}.

The theory of the electron transport in mesoscopic microconstrictions (see,
{\it e.g. }\cite{Imry}) explains the conductance quantization, as a result
of the existence of discrete transverse electron states (modes). On
increasing the contact diameter new modes open up and the conductance
increases in a sequence of steps of the height $G_{0}.$ At a finite
voltages, as a result of the splitting of the Fermi surface in the
constriction by the applied bias $eV$ \cite{KOSh} (Fig. 1), step-like
structure of nonlinear conductance occurs at integer multiples $G_{0}/2,$ as
function of the constriction width \cite{GKh}. The new period of quantum
steps is caused by the difference of the maximum energy of electrons with
the different directions of the electron velocity $v_{\parallel }$ along the
contact axis. With increasing the contact width a new quantum mode opens up
not simultaneously for electrons with the energy $\varepsilon _{F}+\frac{eV}{%
2}$ and the energy $\varepsilon _{F}-\frac{eV}{2}.$ Each time, when a
quantum mode opens up for one of the two directions of the vector $%
v_{\parallel }\lessgtr 0,$ the conductance increases on $G_{0}/2$. If the
bias $eV$ is larger than the distances between the energy levels of quantum
modes, it is possible to change the number of opened modes by changing the
voltage $V.$ In this case the conductance jumps in sequence of steps of the
height $G_{0}/2,$ as the function of the voltage $V$. This effect can be
used for a spectroscopy of energy levels in the quantum constrictions \cite
{Zagoskin}. In the theoretical papers the conductance quantization in the 3D
point contacts of the metal with a spherical Fermi surface was considered
\cite{3Dtheor1}, \cite{3Dtheor2}. It was founded that for sufficiently long
constrictions conductance has the step-like dependence on the contact
diameter. For the symmetric model of the contact, because the degeneration
of the electron energy on one of discrete quantum numbers, conductance has
not only steps $G_{0},$ but also steps $2G_{0}$ \cite{3Dtheor1,3Dtheor2}.

In a majority of real metals the Fermi surface is the complicated periodic
surface, which continuously passes through the whole inverse lattices (open
Fermi surface). The energy $\varepsilon _{\parallel }$of the electron motion
in the direction, in which the Fermi surface is opened, is limited $\left(
0\leq \varepsilon _{\parallel }\leq \varepsilon _{1}\right) $ and its
maximal value $\varepsilon _{1}$ may be considerably smaller then Fermi
energy $\varepsilon _{F}.$ That leads to phenomena, such, for example, as
linear magnetoresistance of polycrystals \cite{LifPesch} (Kapitsa effect
\cite{Kapitsa}) or the oscillation of the resistance of monocrystals as a
function of the direction of the strong magnetic field \cite{Pesch}, which
are absent in the conductors with a closed Fermi surface. The most important
the limitation of the electron velocity in some direction in the ''organic
layered metals'', which Fermi surface is the weekly ''warped'' cylinder\cite
{KKP}.

In this paper we analyze the conductance of 3D quantum microconstriction in
a metal with an open Fermi surface. It is shown, that the conductance may
display not only steps $G_{0}$, but also negative steps $-G_{0},$ as a
result of the limited width of quantum conducting subbands $\Delta
\varepsilon =\varepsilon _{\max }-\varepsilon _{\min }$. The point contact
spectrum $\left( dG/dV\right) $ contains two series of maxima. On of them
corresponds to the voltages $eV=\pm 2\left( \varepsilon _{F}-\varepsilon
_{\min }\right) ,$ as in 2D microconstrictions and 3D point contacts in
metals with an isotropic Fermi surface \cite{Zagoskin}. The second series of
maxima satisfies to the condition $eV=\pm 2\left( \varepsilon
_{F}-\varepsilon _{\max }\right) $.

If the contact axis coincides with the axis of the open Fermi surface, for
the participation of the $n$th quantum mode in the electrical current, not
only minimal energy of the transverse mode $\varepsilon _{\min }\left( {\bf n%
}\right) $ must be smaller then $\varepsilon _{F}\pm \frac{eV}{2}$ , but $%
\varepsilon _{\max }\left( {\bf n}\right) $ $\geq $ $\varepsilon _{F}\pm
\frac{eV}{2},$ where ${\bf n}=\left( n_{1},n_{2}\right) $ is the set of two
transverse discrete quantum numbers, $\varepsilon _{\min }\left( {\bf n}%
\right) $ and $\varepsilon _{\max }\left( {\bf n}\right) $ are the minimal
and maximal energies of quantum subband, which is characterized by the set $%
{\bf n.}$ As a result with increasing of the contact diameter $d,$ starting
from the energy $\varepsilon _{\max }\left( {\bf n},d\right) =\varepsilon
_{F}\pm \frac{eV}{2},$ the ${\bf n}$th mode does not contribute to the
conductivity.

We consider the model of the microconstriction in the form of a long
ballistic channel of the length $L$ and the diameter $d\ll L,$ which is
smoothly (adiabatically \cite{Glazman}) connected with the bulk metallic
reservoirs. In the long $\left( L\gg d\right) $ ballistic channel the
''duplicated'' electron distribution function $f\left( \varepsilon \right) $
has the form \cite{KOSh,KulYan}
\begin{equation}
f\left( \varepsilon \right) =f_{F}\left( \varepsilon +\frac{eV}{2}sign\left(
v_{z}\right) \right) ,
\end{equation}
where $f_{F}\left( \varepsilon \right) $ is the equilibrium Fermi
distribution. The distribution function $f\left( \varepsilon \right) $ (1)
is valid, if the bias is small $eV\ll \sqrt{\varepsilon _{F}\delta
\varepsilon }$ (where $\delta \varepsilon $ is the characteristic distance
between quantum levels of the transverse motion). With this inequality,
which we supposed is fulfilled, the distribution (1) satisfies to the
condition of the electroneutrality and the electrical field inside the
channel is negligibly small.

The total current flowing through the contact can be described by
Landauer-type formula \cite{Landauer}, which at finite voltages is given by

\begin{equation}
J=\frac{2e}{h}\sum_{{\bf n}}\int_{\varepsilon _{\beta \min }}^{\varepsilon
_{\beta \max }}d\varepsilon \left[ f_{F}\left( \varepsilon -\frac{eV}{2}%
\right) -f_{F}\left( \varepsilon +\frac{eV}{2}\right) \right] ,
\end{equation}
The expression (2) has the clear physical meaning: The applied to the
contact bias $eV$ splits the Fermi surface of the injected to the channel
electrons into two parts ($v_{\parallel }>0$ and $v_{\parallel }<0$) with
maximum energies differing by $eV.$ The net current inside the contact is
determined \ by the contributions of these two electronic beams moving in
opposite directions with energies differing by bias energy $eV.$

After the integration in Eq. (2) over the energy $\varepsilon $ we obtain
the following equation for the ballistic conductance:
\begin{eqnarray}
G &=&\frac{dJ}{dV}=\frac{1}{2}G_{0}\sum_{{\bf n}}\left[ f_{F}\left(
\varepsilon _{\min }+\frac{eV}{2}\right) +f_{F}\left( \varepsilon _{\min }-%
\frac{eV}{2}\right) \right. \\
&&\left. -f_{F}\left( \varepsilon _{\max }+\frac{eV}{2}\right) -f_{F}\left(
\varepsilon _{\max }-\frac{eV}{2}\right) \right] ,\quad  \nonumber
\end{eqnarray}
At $V\rightarrow 0$ formula (3) describes the $G_{0}$ steps of the
conductance as the function of the contact size.

\begin{equation}
G=G_{0}\sum_{{\bf n}}\left[ f_{F}\left( \varepsilon _{\min }\right)
-f_{F}\left( \varepsilon _{\max }\right) \right] ,\quad
\end{equation}

Let us consider the ''model metal'' with the Fermi surface

\begin{equation}
\varepsilon ({\bf p)=}\varepsilon _{0}\left( {\bf p}_{\perp }\right)
+\varepsilon _{1}\left( {\bf p}_{\perp }\right) \cos \left( \frac{%
p_{\parallel }a}{\hbar }\right) ;\quad \varepsilon _{1}<\varepsilon _{0}.
\end{equation}
where $a$ is the separation between the atoms. The ''warped'' cylinder $%
\varepsilon ({\bf p)}$ is an infinite surface in the direction $p_{\parallel
}$ and passes in this direction through all cell of the reciprocal space. If
the contact axis is parallel to the $p_{\parallel }$ axis, the transverse
part $\varepsilon _{0}$ of the total energy is quantized $\varepsilon _{0}=$
$\varepsilon _{0}\left( {\bf n}\right) .$ But in the difference from the
spherical Fermi surface, the widths of quasi-one-dimensional subbands have
the limited value $\varepsilon _{1}\left( {\bf n}\right) .$ So, if the
energy $\varepsilon _{\max }\left( {\bf n}\right) =\varepsilon _{0}\left(
{\bf n}\right) +\varepsilon _{1}\left( {\bf n}\right) $ is smaller then
Fermi energy $\varepsilon _{F},$ the subband under Fermi level is completely
filled and does not participate in the current. That results in the negative
steps $-G_{0}$ under the condition $\varepsilon _{\max }\left( {\bf n}%
\right) =\varepsilon _{F}.$

Changing the voltage $eV$ we can change the number of opened quantum modes
for different directions of electron velocity \cite{Zagoskin}. In the metal
with the closed Fermi surface, if the bias is larger than distances between
the energy levels, with increasing of the voltage, the number of quantum
modes under the energy level $\varepsilon _{F}+\frac{eV}{2}$ increases (and
each time conductance increases on $\frac{1}{2}G_{0}$ ), but the number of
modes under the level $\varepsilon _{F}-\frac{eV}{2}$ decreases that leads
to the jumps $-\frac{1}{2}G_{0}.$ The peaks on the point-contact spectrum $%
dG/dV$ are determined by the minimal energies of the transverse electron
modes $\varepsilon _{\min }.$\cite{Zagoskin} In the case of an open Fermi
surface the increasing in the bias determines not only this, but the
opposite processes: at some voltages the maximal energies $\varepsilon
_{\max }$ of subbands go trough the energy levels $\varepsilon _{F}\pm \frac{%
eV}{2}$ changing the conductance by value $\pm \frac{1}{2}G_{0}.$ As a
result the point-contact spectrum has two series of sharp peaks at energies $%
\varepsilon _{F}\pm \frac{eV}{2}=\varepsilon _{\min }\left( {\bf n}\right) $
and $\varepsilon _{F}\pm \frac{eV}{2}=\varepsilon _{\max }\left( {\bf n}%
\right) .$ The measuring of the distances between these peaks makes it
possible to find not only the (minimal) energy of quantum modes in the
constriction, but also the width of quantum subbands and its dependence on
the number of the mode.

The Fig.1 illustrates the conductance of the channel with the square
cross-section as a function of the size $d.$ In Figs.2, 3 the voltage
dependence of the quantized conductance and the point contact spectrum of
the same constriction are shown. For the simplicity we used the model of the
Fermi surface, in which $\varepsilon _{0}={\bf p}_{\perp }^{2}/2m$ and $%
\varepsilon _{1}=const.$

{\it \ }In the quasiclassical case, we can use the Poison formula for the
summation over discrete quantum numbers in Eq.(3). Using the method, which
was developed by Lifshits and Kosevich \cite{LifKos}, at zero temperature we
can write the conductance in the form:
\begin{eqnarray}
G &=&G_{Sh}+G_{0}\frac{2}{\pi }\sum_{{\bf k,}i}\sum_{\alpha =1}^{2}\left(
-1\right) ^{\alpha }\left[ \left| {\bf k}\right| ^{1/2}\left| \nabla
\varepsilon \left( {\bf n}_{i,\alpha }\right) \right| \left| K_{i,\alpha
}\right| ^{1/2}\left( {\bf k}\frac{\partial {\bf n}_{i,\alpha }}{\partial
\varepsilon }\right) \right] ^{-1}\cdot \\
&&\sin \left( 2\pi {\bf kn}_{i,\alpha }\pm \frac{\pi }{4}\right) \cos \pi
{\bf k}\frac{\partial {\bf n}_{i,\alpha }}{\partial \varepsilon }eV
\nonumber
\end{eqnarray}
Here $G_{Sh}$ is the Sharvin conductance \cite{KOSh}; $S$ is the square of
the contact; the vector ${\bf k}$ is the aggregate of two positive integer
numbers $k_{1}$ and $k_{2};$ ${\bf n}_{i,1}$ are the coordinates of the
points on the curve $\varepsilon _{\max }\left( {\bf n}\right) =\varepsilon
_{F}$ and ${\bf n}_{i,2}$ are the coordinates of the points on the curve $%
\varepsilon _{\min }\left( {\bf n}\right) =\varepsilon _{F},$ in which the
normal to these curves is parallel to the vector ${\bf k;}$ $K_{i,\alpha }$
is the curvature of the curve $\varepsilon \left( {\bf n}\right)
=\varepsilon _{F}$ in the points ${\bf n}_{i,\alpha }$. The sign before the
phase $\frac{\pi }{4}$ is minus, if in the point ${\bf n}_{i,\alpha }$ the
convexification of the curve is directed in the direction of vector ${\bf k.}
$ In the opposite case, the sign before $\frac{\pi }{4}$ is plus. In the
Eq.(6) summation is over ${\bf k\neq }0$ and all points ${\bf n}_{i,\alpha }$
in the first quadrant. Hence, in the quasiclassical region the conductance
oscillate as a function of the constriction diameter and the applied bias,
and also depends on the maximal energy of electron motion along the
constriction,

Thus, the conductance of three-dimensional point contacts between the metals
with an open Fermi surface may display the positive and negative steps $%
2e^{2}/h,$ as a function of the contact diameter. The negative steps can be
observed in the geometry of an experiment, when the contact axis is parallel
to the direction, in which a Fermi surface is opened. The decreasing of the
conductance is the result of the complete population of quantum subbands
under the Fermi level. The slope of quantum conductance plateaus in $Al$ and
$Pt,$ which have the opened Fermi surface, could be a result of this effect.
Of course, the electronic structure of $Al$ and $Pt$ is very complicated,
and electrical properties of these metals can not be described by using the
simple model (5). The effect of opened part of the Fermi surface may be
masked by the influence to the conductivity of another parts and instead of
negative jumps the negative slope of conductance plateaus was observed in
experiments \cite{Krans}. Recently the study of nonlinear quantum
conductance has been started \cite{Agrait,Rutenbeek}. The ultrasmall
size of a point contact makes it possible produce the bias right up to $1V$
\cite{Agrait} that opens the possibility of the point-contact spectroscopy
of quantum energy modes in three-dimensional contacts. An experimental
investigation of point-contact spectra for different directions of the
contact axis respect to the crystallographical orientation of the sample to
be studied could enable to observe the effects, which have been
theoretically discussed in this paper, in the voltage dependence of the
conductance of quantum microconstrictions.\newpage

{\bf Figure captions.}

Fig.1. Quantum steps of the conductance in the limit $V\rightarrow 0.$ The
solid line - $\varepsilon _{1}=0.9\varepsilon _{F},$ the dashed line $%
\varepsilon _{1}=0.55\varepsilon _{F};$ $T=0.001\varepsilon _{F}.$

Fig.2. The dependence of the conductance on the applied voltage. The solid
line - $\varepsilon _{1}=0.9\varepsilon _{F},$ the dashed line $\varepsilon
_{1}=0.5\varepsilon _{F};$ $T=0.001\varepsilon _{F};$ $d=1.95\lambda _{F}.$

Fig.3. The point contact spectrum of the microconstriction. The solid line -
$\varepsilon _{1}=0.9\varepsilon _{F},$ the dashed line $\varepsilon
_{1}=0.5\varepsilon _{F};$ $T=0.005\varepsilon _{F};$ $d=1.95\lambda _{F}.$

\bigskip

\bigskip

\bigskip

\bigskip

\ \ \ \ \ \ \ \ \ \ \ \ \ \ \ \ \ \ \ \ \ \ \ \ \ \ \ \ \ \ \ \ \ \ \ \ \ \
\ \ \ \ \ \ \ \ \ \ \ \ \ \ \ \ \


\begin{references}
\bibitem{LifKos}  I.M. Lifshits, A.M.Kosevich, Izv. AN SSSR, Ser. Fiz., {\bf %
19, }395 (1955).

\bibitem{Komnik}  Yu.F. Komnik, Physics of thin films.

\bibitem{2Dfirst}  B.J. van Wees, H. van Houten, C.W.J. Beenakker, J.G.
Williamson, L.P. Kouwenhoven, D. van der Marel and C.T. Foxon, Phys. Rev.
Lett. {\bf 60, }848 (1988).

\bibitem{2Dfirst-too}  D.A. Wharam, M. Pepper, H. Ahmed, J.E.F. Frost, D.G.
Hasko, D.C. Peacock, D.A. Ritchie, G.A.C. Jones, J. Phys. C: Solid State
Phys. {\bf 21, }L209 (1988).

\bibitem{STM1}  J.I. Pascual, J. Mendez, J.Gomez-Herrero, A.M. Baro,
N.Garsia, Vu Thien Binh, Phys. Rev. Lett. {\bf 71, }1852 (1993).

\bibitem{STM2}  N. Agrait, J.G. Rodrigo, S. Viera, {\bf \ }Phys. Rev. B {\bf %
47, }12345 (1993).

\bibitem{BJ1}  C.J. Muller, J.M. van Ruitenbeek, L.J. de Jongh, Phys. Rev.
Lett. {\bf 69, }140 (1992).

\bibitem{BJ2}  J.M. Krans, C.J. Muller, I.K. Yanson, Th.C.M. Govaert, R.
Hesper, J.M. van Ruitenbeek, Phys. Rev. B {\bf 48, }14721 (1993).

\bibitem{Krans}  J.M. Krans, C.J. Muller, J.M. van Ruitenbeek, Physica B,
{\bf 194-196,} 1033 (1994).

\bibitem{Krans2}  J.M. Krans, J.M. van Ruitenbeek, V.V.Fisun, I.K. Yanson
,L.J. de Jongh, Nature, {\bf 375,} 767 (1995){\bf \ }

\bibitem{Imry}  Y. Imry, Introduction to Mesoscopic Physics, Oxford, New
York (1997).

\bibitem{KOSh}  I.O. Kulik, A.N. Omelyanchouk, R.I. Shekhter, Sov. J. Low
Temp. Phys. {\bf 7}, 740 (1977).

\bibitem{KulYan}  I.O. Kulik, I.K. Yanson, Sov. J. Low Temp. Phys., {\bf 4, }%
596 (1978).

\bibitem{GKh}  L.I. Glazman, A.V. Khaetskii, JETP Lett., {\bf 48, }591
(1988).

\bibitem{Landauer}  R. Landauer, IBM J. Res. Dev., {\bf 1, }233 (1957);
Phil. Mag., {\bf 21,} 863 (1970).

\bibitem{Zagoskin}  A.M. Zagoskin, JETP Lett., {\bf 52,} 435 (1990).

\bibitem{Glazman}  L.I. Glazman, G.B. Lesovik, D.E. Khmel'nitskii, R.I.
Shekhter, JETP Lett. {\bf 48, }238 (1988).

\bibitem{2Dtheor1}  A. Szafer, A.D. Stone, Phys. Rev. Lett. {\bf 62, }300
(1989).

\bibitem{3Dtheor1}  E.N.Bogachek, A.M. Zagoskin, I.O. Kulik, Sov. J. Low
Temp. Phys. {\bf 16}, 796 (1990).

\bibitem{3Dtheor2}  J.A. Torres, J.A. Pascual and J.J. Saenz, Phys. Rev. B
{\bf 49, }16581 (1994).

\bibitem{LifPesch}  I.M. Lifshits, V.G. Peschansky, Sov. Phys. JETP, {\bf 8,
}875 (1958).

\bibitem{Kapitsa}  P.L. Kapitsa, Proc. Roy. Soc., {\bf A129, }358 (1928).

\bibitem{Pesch}  V.G. Peschansky, J.A. Roldan Lopes, Toji Gnado Jao, Journ.
de Physique, {\bf 1, }1469 (1991).

\bibitem{KKP}  O.V. Kirichenko, Yu.A. Kolesnichenko, V.G. Peschansky,
Electron Phenomena in Layered Conductors, Physics Reviews (Edit.
I.M.Khalatnikov), {\bf 18, }Part 4 (1998).

\bibitem{Agrait}  C. Untiedt, G. Rubo Bollinger, S. Vieira, N. Agrait, \ In
Quantum and Mesoscopic Phenomena and Mesoscopic Devices in Microelectronics,
Bilkent University, Ankara, Terkey, 78 (1999).

\bibitem{Rutenbeek}  B.Ludoph, M.H. Devoret, D.Esteve, C.Urbina and J.M. van
Ruitenbeek, Phys. Rev. Lett. {\bf 82, }1530 (1999).
\end{references}
\end{document}